\documentclass[aps,amsmath,graphicx,11pt,titlepage]{revtex4}
\usepackage[dvips]{color,graphicx}
\usepackage{amsmath}
\usepackage{amssymb}
\usepackage{braket}

\usepackage[english]{babel}

\def\slasha#1{\setbox0=\hbox{$#1$}#1\hskip-\wd0\hbox to\wd0{\hss\sl/\/\hss}}
\def\slashb#1{\setbox0=\hbox{$#1$}#1\hskip-\wd0\dimen0=5pt\advance
       \dimen0 by-\ht0\advance\dimen0 by\dp0\lower0.5\dimen0\hbox
         to\wd0{\hss\sl/\/\hss}}

\newcommand {\E}[1]{\times 10^{#1}}	
\newcommand {\e}[1]{\mathrm{~#1}}       

\newcommand{\mc}[1]{\mathcal{#1}}

\begin{document}

\begin{titlepage}

\vspace{.5cm}

\begin{center}


{\huge \bf Updated constraints on new physics  }

\vspace{0.3cm}

{\huge \bf  in rare charm decays }

\vspace{.7cm}

{\large  Svjetlana Fajfer, Nejc Ko\v{s}nik and  Sa\v{s}a Prelov\v{s}ek \\}

\vspace{0.5cm}

{\it Department of Physics, University of Ljubljana, 

Jadranska 19, 1000 Ljubljana, Slovenia}

\vspace{0.1cm}

and

\vspace{0.1cm}

{\it J. Stefan Institute, Jamova 39, P. O. Box 300, 1001 Ljubljana
}

\end{center}

\centerline{\large \bf ABSTRACT}
Motivated by recent experimental results on charm physics we
investigate implications of the updated constraints of new physics in
rare charm meson decays. We first reconsider effects of the
MSSM in $c\to u \gamma$ constrained by the recent experimental
evidence on $\Delta m_D$ and find, that due to the dominance of long
distance physics, $D \to V \gamma$ decay rates cannot be modified by
MSSM contributions.  Then we consider effects of the extra heavy up
vector-like quark models on the decay spectrum of $D^+ \to \pi^+
\ell^+ \ell^-$ and $D_s^+ \to K^+ \ell^+ \ell^-$ decays. We find a
possibility for the tiny increase of the differential decay rate in
the region of large dilepton mass.  The $R$-parity violating
supersymmetric model can also modify short distance dynamics in $c \to
u \ell^+ \ell^-$ decays.  We constrain relevant parameters using current
upper bound on the $D^+ \to \pi^+ \ell^+ \ell^-$ decay rate and
investigate impact of that constraint on the $D_s^+ \to K^+ \ell^+
\ell^-$ differential decay dilepton distribution. Present bounds still
allow small modification of the standard model differential decay rate
distribution.

\vspace{0.5cm}

\end{titlepage}

\section{ Introduction}
Recently the evidence for $D^0 - \bar{D}^0$ oscillations has been
reported by Belle and BaBar collaborations \cite{Belle0,BaBar0}.
 Combining the measured quantities~\cite{schwartz} indicates
 nonzero $\Delta \Gamma$ as well as nonzero $\Delta m_D$ 
\begin{equation}
\label{delta_md}
x=\Delta m_D/\Gamma_D=0.0087 \pm 0.003~. 
\end{equation}
These results immediately stimulated many studies (see eg.
\cite{Nir0,IT0,Buras0,chen,Li,Golowich}
).  
The obtained results for the relevant
parameters describing $D^0 - \bar{D}^0$ mixing are not in favor of
new physics effects. However, they give additional constraints
on physics beyond the standard model~(SM) as already given in papers
\cite{IT0,Buras0}.  On the other hand, the study of rare $D$ meson
decays is not considered to be very informative in current searches of
physics beyond the standard model
\cite{burdman1,burdman2,burdman3,FP-LH,prelovsek1,prelovsek2,prelovsek3,bigi0},
as it is expected from ``b'' physics.  Namely, most of the charm meson
processes, where the flavor changing neutral currents~(FCNC) effects might be
present like $c \to u$ and $c \bar u \leftrightarrow \bar c u$
transitions, are dominated by the standard model long-distance
contributions~\cite{burdman1,burdman2,burdman3,FP-LH,prelovsek0,prelovsek1,prelovsek2,prelovsek3,bigi0}.

Due to the GIM mechanism and smallness of the down-type quark masses
the radiative $c \to u \gamma$ decay rate is strongly suppressed at
the leading order in the SM \cite{burdman1,prelovsek1}.  The QCD
effects enhance it up to the order of $10^{-8}$~\cite{greub}.
New bounds on the mass insertion parameters within minimal
supersymmetric SM~(MSSM)~\cite{Nir0,IT0} are derived using the $D^0 -
\bar{D}^0$ oscillations.  We include into consideration the possible
effect of MSSM with non-universal soft breaking terms on $c \to u
\gamma$ along the lines of \cite{masiero,PW}.
Although this approach leads to the enhancement of the SM value by a
factor 10, it is too small to give any observable effects in $D \to
V\gamma$ decays ($V$ is a light vector meson).
The dominating long-distance~(LD) contributions in the $ D \to
V\gamma$ decays give the branching ratios of the order
$\mathrm{Br}\sim 10^{-6}$~\cite{burdman1,prelovsek1}, which makes the
search for new physics effects impossible.

Another possibility to search for the effects of new physics in the
charm sector is offered in the studies of $D \to X \ell^+ \ell^-$
decays which might be result of the $c \to u \ell^+ \ell^-$ FCNC
transition~\cite{chen,burdman2,burdman3,prelovsek0,prelovsek2,prelovsek3}.
Here $X$ can be light vector meson $V$ or pseudoscalar meson $P$.
Within SM inclusion of renormalization group improved QCD corrections
for the inclusive $c\to u \ell^+\ell^-$ gave an additional significant
suppression leading to the rates $\Gamma(c\to
ue^+e^-)/\Gamma_{D^0}=2.4\times 10^{-10}$ and $\Gamma(c\to
u\mu^+\mu^-)/\Gamma_{D^0}=0.5\times 10^{-10}$~\cite{jure}. These
transitions are largely driven by a virtual photon at low dilepton
mass $m_{\ell\ell}\equiv \sqrt{(p_+ + p_-)^2}$, while the total rate
for $D \to X \ell^+ \ell^-$ is dominated by the LD resonant
contributions at dilepton masses
$m_{\ell\ell}=m_\rho,m_\omega,m_\phi$~\cite{burdman2, prelovsek3}.

New physics could possibly modify the dilepton mass distribution below
$\rho$ or distribution above $\phi$ resonance.  In the case of
$D\to\pi \ell^+\ell^-$ there is a broad kinematical region of dilepton mass
above $\phi$ resonance which presents an unique possibility to study
$c\to u\ell^+\ell^-$ at high $m_{\ell\ell}$~\cite{prelovsek3}.

The leading contribution to $c\to u\ell^+\ell^-$ in general MSSM with
the conserved $R$-parity comes from one-loop diagram with gluino and
squarks in the loop~\cite{burdman2,prelovsek3,sasa}.  It proceeds via
virtual photon and enhances the $c\to u\ell^+\ell^-$ spectrum at small
$m_{\ell\ell}$.  We find that bounds on the mass insertion parameters
make the abovementioned enhancement in $D$ rare
decay~\cite{burdman2,burdman3} to be negligible.

Some models of new physics contain an extra up-like heavy quark
singlet~\cite{faj-ichep06} inducing the FCNCs at tree level for the
up-quark sector~\cite{FP-LH,barger,lang,abel,higuchi}, while the
neutral current for the down-like quarks is the same as in the SM.
The stringest bound on these models comes from the recent bound on
$\Delta m $ in the $D^0 - \bar{D}^0$ transition as given in
\cite{Belle0,BaBar0}. In our calculation, we analyze how these bounds
on the FCNC vertex $cuZ$ affect the $D \to P \ell^+ \ell^-$ decays.  A
particular version of the model with tree-level up-quark FCNC
transitions is the Littlest Higgs model \cite{lee}.  In this case the
magnitude of the relevant $c \to u Z$ coupling is even further
constrained by the large scale $f\geq {\cal O}(1~{\rm TeV})$ using the
precision electroweak data. The smallness implies that the effect of
this particular model on $c\to u\ell^+\ell^-$ decay and relevant rare
$D$ decays is insignificant \cite{FP-LH}.

Among discussed models of new physics the supersymmetric extension of
the SM including the $R$-parity violation is still not constrained as
other new physics models.  As noticed by \cite{burdman2,jure} one can
test some combinations of the $R$-parity violating contributions in
$D^+ \to \pi^+ \ell^+ \ell^-$ decays.  We place new constraints on the
relevant parameters and search for the effects of new physics in the
$D_s^+ \to K^+ \ell^+ \ell^-$ decays which might be interesting for
the experimental studies.

There are intensive experimental efforts by
CLEO~\cite{CLEO_pll,CLEO_vll} and
FERMILAB~\cite{FERMILAB_pll,FERMILAB_vll} collaborations to improve
the upper limits on the rates for $D\to X \ell^+\ell^-$ decays.  Two
events in the channel $D^+\to \pi^+e^+e^-$ with $m_{ee}$ close to
$m_\phi$ have already been observed by CLEO~\cite{CLEO_pll}. The other
rare $D$ meson decays are not so easily accessible by experimental
searches, but with the plans to make more experimental studies in rare
charm decays at CLEO-c, Tevatron and at charm physics sections at
present $B$-factories and in the future at LHC-b facilities makes the
study of rare $D$ decays more attractive.  

In order to compare effects of new physics and the standard model we
have to determine size of the LD contributions.  As in the case of
$D^+\to \pi^+e^+e^-$ decay we use experimental data on the $D_s$
nonleptonic decays accompanied by the vector meson dominance as we
did in~\cite{FP-LH}.

The paper is organized as follows. In Section~2 we consider the impact
of new charm mixing bounds on the $c \to u \gamma$ decay. In Sec.~3 we
study how new physics affects $c \to u \ell^+ \ell^-$ and $D \to P
\ell^+ \ell^-$ decays. We present framework for calculating SD effects
as well as the details of LD calculations. In Sec.~4 we discuss our
results and in Sec.~5 we make a short summary.

\section{$c\to u\gamma$ decay}
Given the recent observation of $D^0-\bar D^0$ mixing, we re-evaluate
the possible effect of MSSM on $c\to u\gamma$. Since the model with
universal soft-breaking terms is known to have negligible
effect~\cite{PW}, we consider the model with non-universal soft
breaking terms. We consider only the gluino exchange diagrams through
$(\delta_{12}^u)_{LR,RL}$ mass insertions, since the remaining SUSY
contributions can not have sizable effect~\cite{masiero,PW}.  The
maximal value of $(\delta_{12}^u)_{LR,RL}$ insertion has been
constrained by saturating $x=\Delta m_D/\Gamma=(4.8\pm 2.8)\E{-3}$
with the gluino exhange in~\cite{IT0}. The results corresponding to
the measured $x=(8.7\pm 3)\times 10^{-3}$~\cite{schwartz,Golowich} are
shown in second column of Table \ref{tab.insertions}.  Another
constraint is obtained by requiring the minima of MSSM scalar
potential do not break electric charge or color and that they are
bounded from above $(\delta_{12}^u)_{LR,RL}\leq \sqrt{3} m_c/m_{\tilde
  q}$~\cite{CD} , with values given in third column of
Table~\ref{tab.insertions}.  The second constraint is obviously
stronger for $m_{\tilde q}\geq 350$ GeV, while $\Delta m_D$ gives more
stringent constraint for lighter squarks.  Using
$(\delta_{12}^u)_{LR,RL}\leq \sqrt{3} ~m_c/m_{\tilde q}$, $m_{\tilde
  q}=m_{\tilde g}=350$ GeV, $m_c=1.25$ GeV and expressions
from~\cite{PW} we get the upper bound
\begin{equation}
\Gamma(c\to u\gamma)/\Gamma_{D^0}\leq 8\times 10^{-7},
\end{equation}
which is one order of magnitude larger than the standard model
prediction \mbox{$\Gamma(c\to
  u\gamma)/\Gamma_{D^0}=2.5\E{-8}$}~\cite{greub}.
\begin{table}[h]
\begin{center}
\begin{tabular}{|c|c|c|}
  \hline
  $m_{\tilde q}=m_{\tilde g}$ & $|(\delta_{12}^u)_{LR,RL}|$ & $|(\delta_{12}^u)_{LR,RL}|$ \\
  & from $\Delta m_D$         & from stability bound      \\
  \hline
  350 GeV & 0.007 & 0.006 \\
  500 GeV & 0.01 & 0.004 \\
  1000 GeV & 0.02  & 0.002\\                 
  \hline
\end{tabular}
\caption{Upper bounds on mass insertions $|(\delta_{12}^u)_{LR,RL}|$ from measured $\Delta m_D$ and stability bound \cite{CD}.}
\label{tab.insertions}
\end{center}
\end{table} 

However, this possible SUSY enhancement by factor $10$ would not
affect the rate of the $D\to V\gamma$ decays, which are completely
dominated by LD contributions with $\mathrm{Br}\sim 10^{-6}$
\cite{burdman1,burdman2,burdman3,prelovsek0,prelovsek1}. The only
window for probing the $c\to u\gamma$ enhancement remains the $B_c\to
B_u^*\gamma$ decay, where LD contributions are strongly
suppressed~\cite{FPS99}.

\section{$c\to u\ell^+\ell^-$ and $D\to P\ell^+\ell^-$ decays}

\subsection{SD Effects}
The $c \to u \ell^+ \ell^-$ transition is driven by the low-energy
effective Lagrangian
\begin{equation} \label{eq:SDLagrangian}
  \mc{L}^{\mathrm{SD}}_{\mathrm{eff}}= \frac{G_F}{\sqrt{2}} V_{cb}^* V_{ub}
  \sum_{i=7,9,10} C_i Q_i,
\end{equation}
given in terms of four-quark operators
\begin{align}
  Q_7 &= \frac{e}{8\pi^2} m_c F_{\mu\nu} \bar{u}
  \sigma^{\mu\nu}(1+\gamma_5) c,\\
  Q_9 &= \frac{e^2}{16\pi^2} \bar{u}_L \gamma_\mu c_L \bar{\ell} \gamma^\mu \ell,\\
  Q_{10} &= \frac{e^2}{16\pi^2} \bar{u}_L \gamma_\mu c_L \bar{\ell}
  \gamma^\mu \gamma_5 \ell.
\end{align}
and the corresponding Wilson coefficients $C_{7,9,10}$. $F_{\mu\nu}$
is the electromagnetic field strength, while $q_L =
\frac{1}{2}(1-\gamma_5)\,q$ are the left-handed quark fields. Wilson
coefficients are taken at the scale $\mu = m_c$.

Since we consider exclusive decay modes $D \to P \ell^+ \ell^-$ we
have to employ form factor description of the four-quark operators
evaluated between two mesonic states. We use the standard
parameterization
\begin{align}
  \Braket{P(k) | \bar{u} \gamma^\mu (1-\gamma_5) c | D(p)} &= (p+k)^\mu f_+(q^2) + (p-k)^\mu f_-(q^2),\\
  \Braket{P(k) | \bar{u} \sigma^{\mu\nu} (1\pm\gamma_5) c | D(p)} &=
  i s(q^2)\left[(p+k)^\mu q^\nu - q^\mu (p+k)^\nu \pm i
    \epsilon^{\mu\nu\alpha\beta} (p+k)_\alpha q_\beta \right],
\end{align}
where $P=\pi^+\,(K^+)$ in the case of $D=D^+\,(D_s^+)$. The momentum
transfer $q=p-k$ is also the momentum of the lepton pair. For the
$f_+$ form factor we use the double pole parameterization of
Ref.~\cite{hep-ph/0412140}
\begin{equation}
f_+(q^2) = \frac{f_+(0)}{(1-x)(1 - a x)},
\end{equation}
where $x=q^2/m_{D^{*}}^2$, $f_+(0) = 0.617$, and $a=0.579$. We
approximate $s(q^2)$ by $f_+(q^2)/m_{D}$, which is valid in
the limit of heavy $c$-quark and zero recoil limit~\cite{IW}.  This
relation can be modified as noticed in~\cite{Becirevic,Jansen}.
However, in our case this modification cannot give significant
effects.  Finally, we arrive at the short distance amplitude for $c
\to u \ell^+ \ell^-$ decay
\begin{equation}
  \label{eq:SDAmpl}
  \mc{A}_{\mathrm{SD}} = -i \frac{4\pi\alpha G_F}{\sqrt{2}} V_{cb}^* V_{ub} \left[
    \left(\frac{C_7}{2\pi^2}\frac{m_c}{m_D} +
      \frac{C_9}{16\pi^2}\right) \bar{u}(p_-) \slasha{p}
    v(p_+) +\frac{C_{10}}{16\pi^2} \bar{u}(p_-)
    \slasha{p} \gamma_5 v(p_+)\right]f_+(q^2).
\end{equation}
$p$, $p_+$ and $p_-$ are the momenta of the initial $D$ meson and the
lepton pair in the final state, respectively. 

\subsubsection{Standard model}
The SM rate is dominated by the photon exchange, where $c \to u
\gamma$ is a two loop diagram induced by effective weak vertex and a
gluon exchange~\cite{greub,pham,jure,FP-LH}. The effective Wilson
coefficient is~\cite{FP-LH}
\begin{equation} 
  V_{cb}^* V_{ub} \hat{C}_7^{\textrm{eff}} = V_{cs}^* V_{us} (0.007 + 0.020i)(1\pm
  0.2).
\end{equation}
The remaining two Wilson coefficients are subdominant in the SM and
we ignore them in further analysis. $C_9$ is small due to the effects of
the renormalization group, while the coefficient $C_{10}$ is
completely negligible in the SM~\cite{FP-LH}.

\subsubsection{Models with extra heavy up vector-like quark singlet}
The class of models with an extra up-like quark singlet~(EQS)
naturally accommodate FCNCs at tree level~\cite{faj-ichep06,FP-LH}
\begin{equation}
  \mc{L}_{NC} = \frac{g}{\cos\theta_W} Z_\mu (J^\mu_{W^3}-\sin^2\theta_W J_{EM}^\mu).
\end{equation}
$J_{EM}^\mu$ is the electromagnetic current, while the weak neutral
current
\begin{equation}
  \label{eq:neutralcurrent}
  J_{W^3}^\mu = \frac{1}{2}\bar{U}_L^m \gamma^\mu \Omega U_L^m
 - \frac{1}{2} \bar{D}_L^m \gamma^\mu D_L^m
\end{equation}
mixes up-type quarks~\cite{lee}, where $U^m$ and $D^m$ are the quark
mass eigenstates. The transition matrix to the mass eigenbasis for
the up-type quarks is $4\times 4$ unitary matrix $T_L^U$, which causes
tree-level FCNCs in the interaction term $J_{W^3}^\mu Z_\mu$ in the up
sector.  The mixing matrix contains only the elements of the last
column of matrix $T_L^U$
\begin{equation}
\Omega=
\begin{pmatrix}
  1-|\Theta_u|^2 & -\Theta_u \Theta_c^* & -\Theta_u \Theta_t^* & -\Theta_u \Theta_T^*\\
  -\Theta_c \Theta_u^* & 1-|\Theta_c|^2 & -\Theta_c \Theta_t^* &
  -\Theta_c \Theta_T^*\\
  -\Theta_t \Theta_u^* & -\Theta_t \Theta_c^* & 1-|\Theta_t|^2 & -\Theta_t \Theta_T^*\\
  -\Theta_T \Theta_u^* & -\Theta_T \Theta_c^* & -\Theta_T \Theta_t^* &
  1-|\Theta_T|^2
\end{pmatrix}.
\end{equation}
The unitarity of the extended CKM matrix then implies that
off-diagonal elements of $\Omega$ are non-zero, e.g. $\Omega_{uc}
\equiv -\Theta_u \Theta_c^* = V_{ud} V_{cd}^* + V_{us} V_{cs}^* +
V_{ub} V_{cb}^* \neq 0.$ The low-energy effective description is
encoded in Wilson coefficients $C_9$ and $C_{10}$. Relative to the
negligible SM values, they are modified by the presence of an extra
up-like quark:
\begin{align}
  V_{ub} V_{cb}^* \delta C_9 &= \frac{4\pi}{\alpha} \Omega_{uc} (4\sin^2\theta_W-1)\\
  V_{ub} V_{cb}^* \delta C_{10} &= \frac{4\pi}{\alpha} \Omega_{uc} ,
\end{align}
The element $\Omega_{uc}$ of the up-type quark mixing matrix is
constrained by the measurements of $D^0-\bar{D^0}$
mixing~\cite{Belle0,BaBar0,Nir0} and using expression $\Delta m_D =
2\E{-7} \left|\Omega_{uc}\right|^2 \e{GeV}$~\cite{lee}:
\begin{equation}
  \Omega_{uc} < 2.8\E{-4}.
\end{equation}

\subsubsection{Minimal supersymmetric SM}
The leading contribution to $c \to u \ell^+ \ell^-$ in general MSSM
with conserved $R$-parity comes from the gluino exchange diagram via
virtual photon and significantly enhances $c\to u\ell^+ \ell^-$ at
small $m_{\ell\ell}$. This MSSM enhancement can not be so drastic in
hadronic decays, since gauge invariance imposes additional factor of
$m_{\ell\ell}^2$ for $D \to P \ell^+ \ell^-$ decays, while $D \to V
\ell^+ \ell^-$ has large long distance contribution at small
$m_{\ell\ell}$ just like $D \to V \gamma$.

In the MSSM with broken $R$-parity (MSSM$\slasha{R}$), the $c \to u
\ell^+ \ell^-$ process is mediated by the tree-level exchange of down
squarks~\cite{burdman2}. Integrating them out leads to the effective
four-quark interaction
\begin{equation}
  \label{eq:RPVLagrangian}
  \mc{L}_{\mathrm{eff}} = \sum_{i,k=1}^3 \frac{\tilde{\lambda}'_{i2k}\tilde{\lambda}'_{i1k}}{2 M^2_{\tilde{d}_R^k}} (\bar{u}_L \gamma^\mu c_L)(\bar{\ell}_L \gamma_\mu \ell_L).
\end{equation}
$\tilde{\lambda}'_{ijk}$ are the CKM-rotated couplings between the
$L$, $Q$ and $D$ supermultiplets in the
superpotential~\cite{burdman2}.  In our
notation~(\ref{eq:SDLagrangian}), the contribution to the Wilson
coefficients is~\cite{jure}
\begin{align}
  V_{cb}^* V_{ub} \delta C_9 = -V_{cb}^* V_{ub} \delta C_{10}= \frac{2
    \sin\theta_W^2}{\alpha^2} \sum_{k=1}^3
  \left(\frac{m_W}{M_{\tilde{d}_R^k}}\right)^2 \tilde{\lambda}_{i2k}'
  \tilde{\lambda}_{i1k}',
\end{align}
where $i=1~(2)$ contributes to the $e^+e^-\,\,(\mu^+\mu^-)$ mode.
The $\tilde{\lambda}_{12k}'$ and $\tilde{\lambda}_{11k}'$ have been
constrained from the charged current
universality~\cite{Allanach,jure}, while the strictest constraint on
$\sum_k \tilde{\lambda}_{22k}'\tilde{\lambda}_{21k}'$
comes~\cite{jure} from the experimental limit $\mathrm{Br}(D^+ \to
\pi^+ \mu^+ \mu^-) = 8.8\E{-6}$~\cite{Dtopimumu}.
We shall reanalyze the latter case, where LD physics
generates the fair amount of experimental branching ratio. It is
sensible to use an approach, where one takes into account the
interference between LD and MSSM$\slasha{R}$ part of the amplitude to constrain
the couplings of the MSSM$\slasha{R}$.

\subsection{Long distance contributions in $D \to P \ell^+ \ell^-$}
Knowledge of the LD contributions is crucial, if we want to isolate
short distance physics in the decays of type $D \to P \ell^+ \ell^-$.
Following procedure described in~\cite{FP-LH} we consider long
distance contributions by employing the resonant decay modes, in which
$D$ first decays to $P$ and a virtual neutral vector meson $V_0$,
followed by decay of $V_0 \to \gamma \to\ell^+ \ell^-$. First stage of
the decay is controlled by effective weak non-leptonic Lagrangian
\begin{equation} \label{eq:LDLagrangian}
  \mc{L}^{\mathrm{LD}}_\mathrm{eff} = -\frac{G_F}{\sqrt{2}} 
  \sum_{q=d,s} V_{uq} V_{cq}^* \left[a_1 \bar{u} \gamma^\mu (1-\gamma_5) q \, 
    \bar{q} \gamma_\mu (1-\gamma_5)c +  a_2 \bar{u} \gamma^\mu (1-\gamma_5) c \, 
    \bar{q} \gamma_\mu (1-\gamma_5)q   \right]
\end{equation}
The effective Wilson coefficients on the scale $m_c$ are~\cite{jure}
\begin{equation}
  a_1 = 1.26,\quad a_2 = -0.49.
\end{equation}
The flavour structure of (\ref{eq:LDLagrangian}) allows $V_0$ to be
either $\rho$, $\omega$ or $\phi$. Since branching ratios of separate
stages in the cascade are well measured, we will not work in a
particular theoretical model, but will instead try to make the best
use of experimental data currently available. Here we follow the
lines of Ref.~\cite{FP-LH}. For a cascade, we write~\cite{lichard}
\begin{equation}
  \frac{\mathrm{d}\Gamma}{\mathrm{d}q^2}(D_s \to K V_0 \to K \ell^+\ell^-)
  =\frac{1}{\pi}\Gamma_{D_s \to K V_0} (q^2) \frac{\sqrt{q^2}}{(m_{V_0}^2-q^2)^2+m_{V_0}^2 \Gamma_{V_0}^2} \Gamma_{V_0 \to \ell^+ \ell^-} (q^2).
\end{equation}
Here $\Gamma_{D_s \to K V_0} (q^2)$ and $\Gamma_{V_0 \to \ell^+ \ell^-}$
denote decay rates if $V_0$ had a mass $\sqrt{q^2}$ and these rates
are known experimentally only at $\sqrt{q^2}=m_{V_0}$. Since the
resonances $V_0 = \rho, \omega, \phi$ are relatively narrow
($\Gamma_{V_0} \ll m_{V_0}$), the following relation approximately
holds
\begin{equation} \label{eq:narrowwidth}
  \mathrm{Br} \left[D \to P V_0 \to P  \ell^- \ell^+\right] = \mathrm{Br} \left[D  \to P  V_0 \right] \times \mathrm{Br}\left[V_0  \to  \ell^- \ell^+ \right].
\end{equation}
The phenomenological amplitude ansatz that reproduces the above
behaviour is then~\cite{FP-LH}
\begin{equation} \label{eq:ansatz}
  \mc{A}^{\mathrm{LD}} \left[D_s (p) \to K (p-q) V_0(q) \to K (p-q) \ell^-(p_-) \ell^+(p_+)\right]
  = 
  e^{i \phi_{V_0}} \frac{a_{V_0}}{q^2-m_{V_0}^2 + i m_{V_0} \Gamma_{V_0}}\, \bar{u}(p_-)\, \slasha{p}\,v(p_+).
\end{equation}
The only assumption we made here is that coefficient $a_{V_0}$ is
independent of $q^2$.  We included the phase $\phi_{V_0}$ explicitly, so
that $a_{V_0}$ is real and positive number.
\subsubsection{$D^+ \to \pi^+ \ell^+ \ell^-$}
For the right side of Eq.~(\ref{eq:narrowwidth}) we use experimental
data (Table~\ref{tab:Ddata}), and use ansatz~(\ref{eq:ansatz}) to extract
unknown parameters $a_{V_0}$: $a_\rho = 2.94\E{-9},\, a_\phi = 4.31\E{-9}$.
\begin{table}[!h]
  \centering\begin{tabular}{c||c|c|c}
    mode & $D^+ \to \pi^+ \rho$ & $D^+ \to \pi^+ \omega$ & $D^+ \to \pi^+ \phi$\\\hline 
    Br$\E{3}$ & $1.07 \pm 0.11$ & $<0.34$ & $6.50 \pm 0.70$ 
  \end{tabular}\\
  \caption{Branching ratios of decays of $D^+$ meson to the
    intermediate resonant states~\cite{pdg}.}
\label{tab:Ddata}
\end{table}
Decay mode $D^+ \to \pi^+ \omega$ has not been measured yet, but we
can relate $a_\omega$ and its phase to the well-measured contribution
of the $\rho$ resonance assuming vector meson dominance as
in~\cite{FP-LH}.  Relative phases and magnitudes of the resonances are
extracted by considering the decay mechanism, controlled by the weak
Lagrangian~(\ref{eq:LDLagrangian}) and electromagnetic coupling of
$V_0$ to photon. Then flavour structure of the resonances determines
relative sizes and phases of resonant amplitudes.  Detailed analysis
has already been done in Ref.~\cite{FP-LH}.  The relative phases of
$\rho$ and $\omega$ contributions are found to be opposite in sign,
while for the ratio of they magnitudes it was found that
$a_\omega/a_\rho = 1/3$.  Also the phases of $\rho$ and $\phi$ are
opposite. Thus the final LD amplitude (up to the phase) becomes
\begin{equation}
  \label{eq:LDAmplitudeD}
  i \mc{M}^{\mathrm{LD}} = \left[a_\rho \left(\frac{1}{q^2-m_{\rho}^2 + i m_{\rho} \Gamma_{\rho}} - \frac{1}{3} \frac{1}{q^2-m_{\omega}^2 + i m_{\omega} \Gamma_{\omega}}\right) - \frac{a_{\phi}}{q^2-m_{\phi}^2 + i m_{\phi} \Gamma_{\phi}} \right]\bar{u}(p_-)\, \slasha{p}\,v(p_+).
\end{equation}

\subsubsection{$D_s^+ \to K^+ \ell^- \ell^+$}
Experimental data is not as rich as in the case of non-strange
charmed meson decays~(Table~\ref{tab:Dsdata}).
\begin{table}[h]
  \centering\begin{tabular}{c||c|c|c}
    mode & $D_s^+ \to K^+ \rho$ & $D_s^+ \to K^+ \omega$ & $D_s^+ \to K^+ \phi$\\\hline
    Br$\times 10^3$ & $2.60 \pm .70$ & $-$ & $<0.50$
  \end{tabular}
  \caption{Branching ratios of $D_s^+$ meson to the intermediate
    resonant state~\cite{pdg}.}
\label{tab:Dsdata}
\end{table}
Contributions of $\rho$ and $\omega$ are related like in the case of
$D^+$ meson, namely $a_\omega/a_\rho = 1/3$ with opposite relative
phase between them. In the same way as for the non-strange decays, we
determine $a_\rho = 6.97\E{-9}$. However, the contribution of the
$\phi$ resonance is only limited from above by experimental data and
we have to rely on a theoretical model. Consequently, the total LD
amplitude is a sum of two terms:
\begin{equation}
  \label{eq:LDAmplitudeDs}
  \mc{A}^{\mathrm{LD}} = a_\rho \left(\frac{1}{q^2-m_{\rho}^2 + i m_{\rho} \Gamma_{\rho}} - \frac{1}{3} \frac{1}{q^2-m_{\omega}^2 + i m_{\omega} \Gamma_{\omega}}\right) \bar{u}(p_-)\, \slasha{p}\,v(p_+) +  \mc{A}^{\mathrm{LD}}_\phi. 
\end{equation}
We calculate the $\phi$ part of (\ref{eq:LDAmplitudeDs}) using the
vector meson dominance (VMD) assumption, where the intermediate $\phi$
contributes by decaying into a virtual photon, which further decays to
the lepton pair. Both $a_1$ and $a_2$ parts of the non-leptonic
Lagrangian (\ref{eq:LDLagrangian}) can generate the flavour quantum
numbers of $\phi$ and $K^+$. The $a_1$ part connects initial $D_s^+$
state to $\phi$ through a charged current $(\bar{s} c)_{V-A}$, while
the $(\bar{u} s)_{V-A}$ creates the $K^+$ out of vacuum. Neutral
currents (the $a_2$ part) do the opposite: $D_s^+ \to K^+$ and $0 \to
\phi$. Utilizing the Feynman rules of (\ref{eq:LDLagrangian}), and the
VMD hypothesis we arrive at the $\phi$ contribution to the LD
amplitude
\begin{equation}
  \label{eq:phicontribution}
  \mc{A}^{\mathrm{LD}}_\phi = i \frac{4\pi\sqrt{2}}{3} G_F V_{us }V_{cs}^* \alpha \frac{g_\phi}{q^2 (q^2-m_\phi^2+i m_\phi \Gamma_\phi)}\left[a_1 m_\phi f_K A_0(m_K^2) + a_2 g_\phi f_+(q^2)  \right] \bar{u}(p_-) \slasha{p} v(p_+).
\end{equation}
However, the phase of $\mc{A}^{\mathrm{LD}}_\phi$ relative to the rest
of the amplitude (\ref{eq:LDAmplitudeDs}) remains to be free. The $P
\to V$ transition $D_s^+ \to \phi$ is described by the form factor
$A_0$, which we take from~\cite{kamenik}. In our calculations we
consider gauge invariant amplitude for $D_s^+ \to K^+ \ell^+ \ell^-$
in which $1/q^2$ dependence is cancelled.

\section{RESULTS}
Using the approach, described in Sec.~3, we analyze impact of short
distance physics on long-distance resonant background.  Since the SD
contribution of SM is completely overshadowed by LD, we will only
consider the EQS and MSSM$\slasha{R}$ models of new physics, to see if
the experimental searches for them are still viable in $D \to X
\ell^+\ell^-$ decays.  Current constraints on EQS model coming from the
$D^0-\bar{D}^0$ mixing already indicate the dominance of LD
contributions in the total decay rate. On the other hand, the
contribution of MSSM$\slasha{R}$ is not as constrained and one should
still see the deviations from the LD contribution away from the
resonant region of the phase space.

We shall analyze the dilepton squared mass $(q^2=m_{\ell\ell}^2)$
distribution of the branching ratios. We fix free phases in the
amplitude in a way which maximizes branching ratio for the considered
decay mode.

\subsection{$D^+ \to \pi^+ \ell^+ \ell^-$}
\begin{figure}[!h]
\begin{tabular}{cc}
  \includegraphics[angle=-90,width=0.5\textwidth]{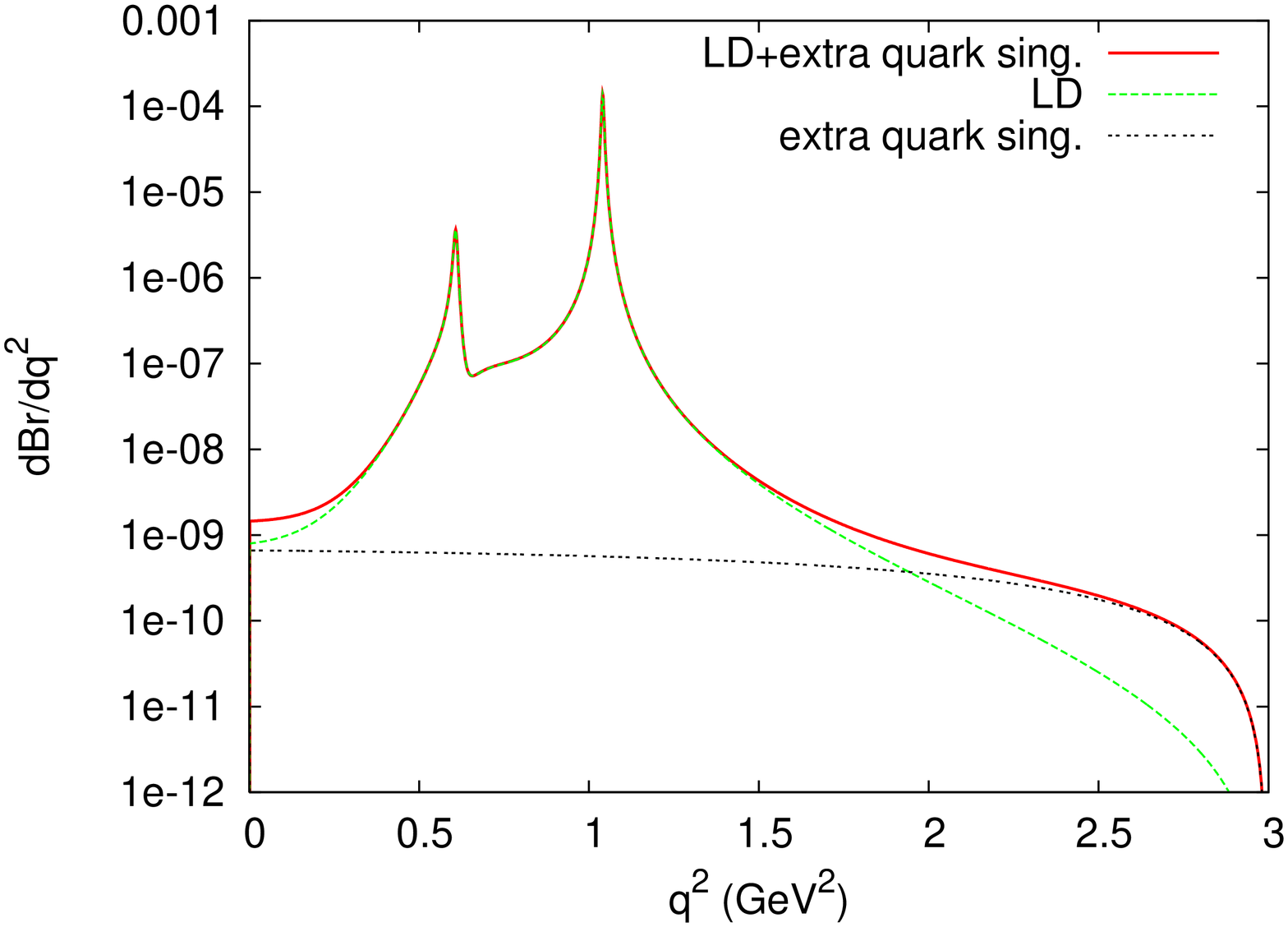} &
  \includegraphics[angle=-90,width=0.5\textwidth]{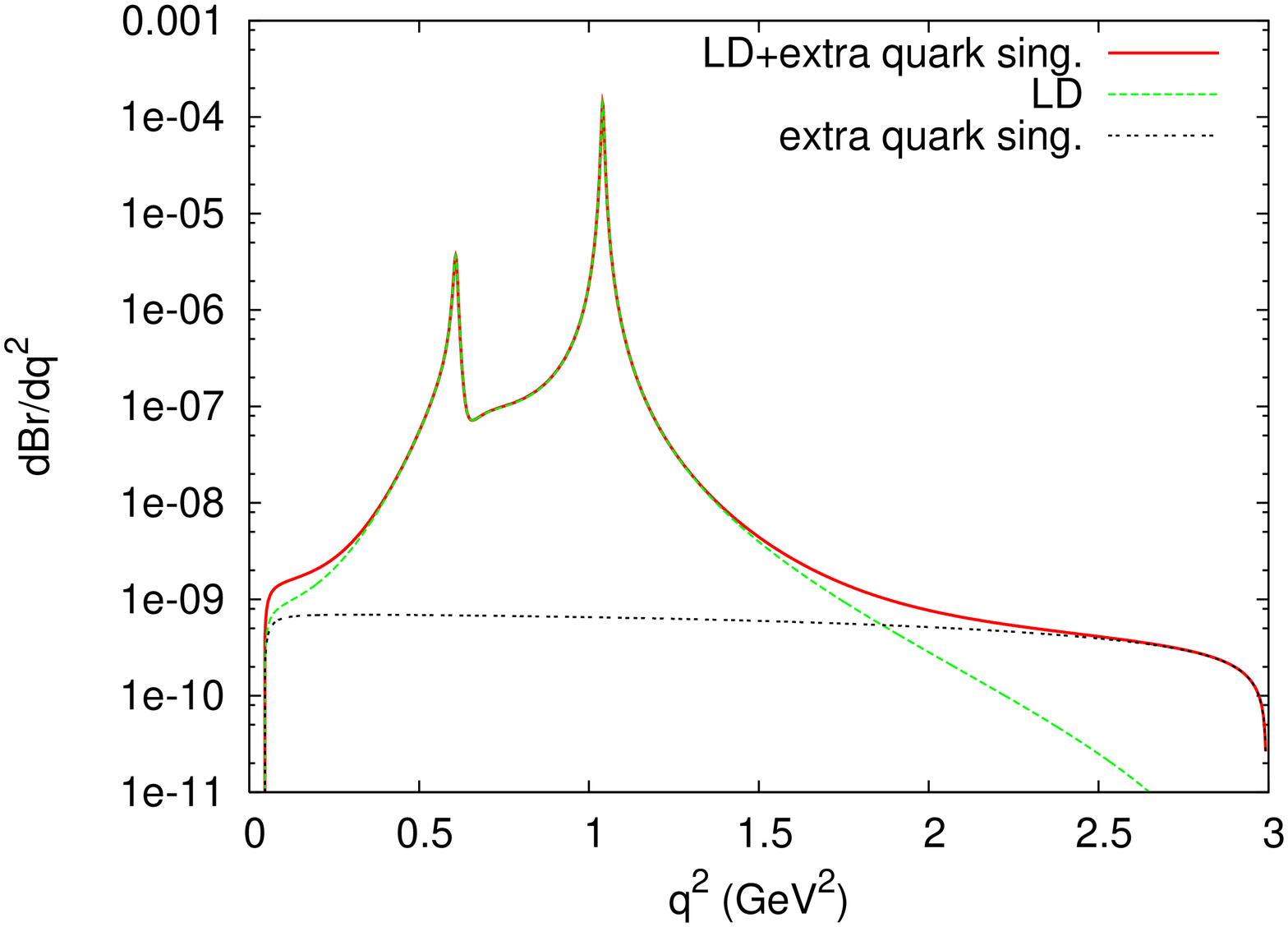}
\end{tabular}
\caption{Distributions of the maximal branching ratios in the model
  with extra quark singlet for the decay modes $D^+ \to \pi^+ e^+
  e^-$~(left) and $D^+ \to \pi^+ \mu^+ \mu^-$~(right). Full line
  represents the combined LD and SD contributions.}
\label{fig:DEQS}
\end{figure}

\begin{table}[h]
\begin{tabular}{c||c||c|c||c|c}
  mode & LD & extra heavy quark & LD+extra heavy quark & MSSM$\slasha{R}$ & LD+MSSM$\slasha{R}$\\\hline\hline
  $D^+ \to \pi^+ e^+ e^-$ & $2.0\E{-6}$ & $1.3\E{-9}$ & $2.0\E{-6}$ & $2.1\E{-7}$ & $2.3\E{-6}$\\\hline
  $D^+ \to \pi^+ \mu^+ \mu^-$ & $2.0\E{-6}$ & $1.6\E{-9}$ & $2.0\E{-6}$ & $6.5\E{-6}$ & $8.8\E{-6}$
\end{tabular}
\caption{Total branching fractions of the $D^+ \to \pi^+ \ell^+ \ell^-$ modes.
  In the first column (LD) are only long-distance BRs. The remaining
  four columns give maximal contributions of the SD physics models 
  alone and also combined contributions of the SD and LD physics.}
\label{tab:Dbr}
\end{table}
Branching fractions are listed in Table~\ref{tab:Dbr}. Clearly, the
EQS model contribution is too small to be observed
(Fig.~\ref{fig:DEQS}).

\begin{figure}[!h]
\begin{tabular}{cc}
\includegraphics[angle=-90,width=0.5\textwidth]{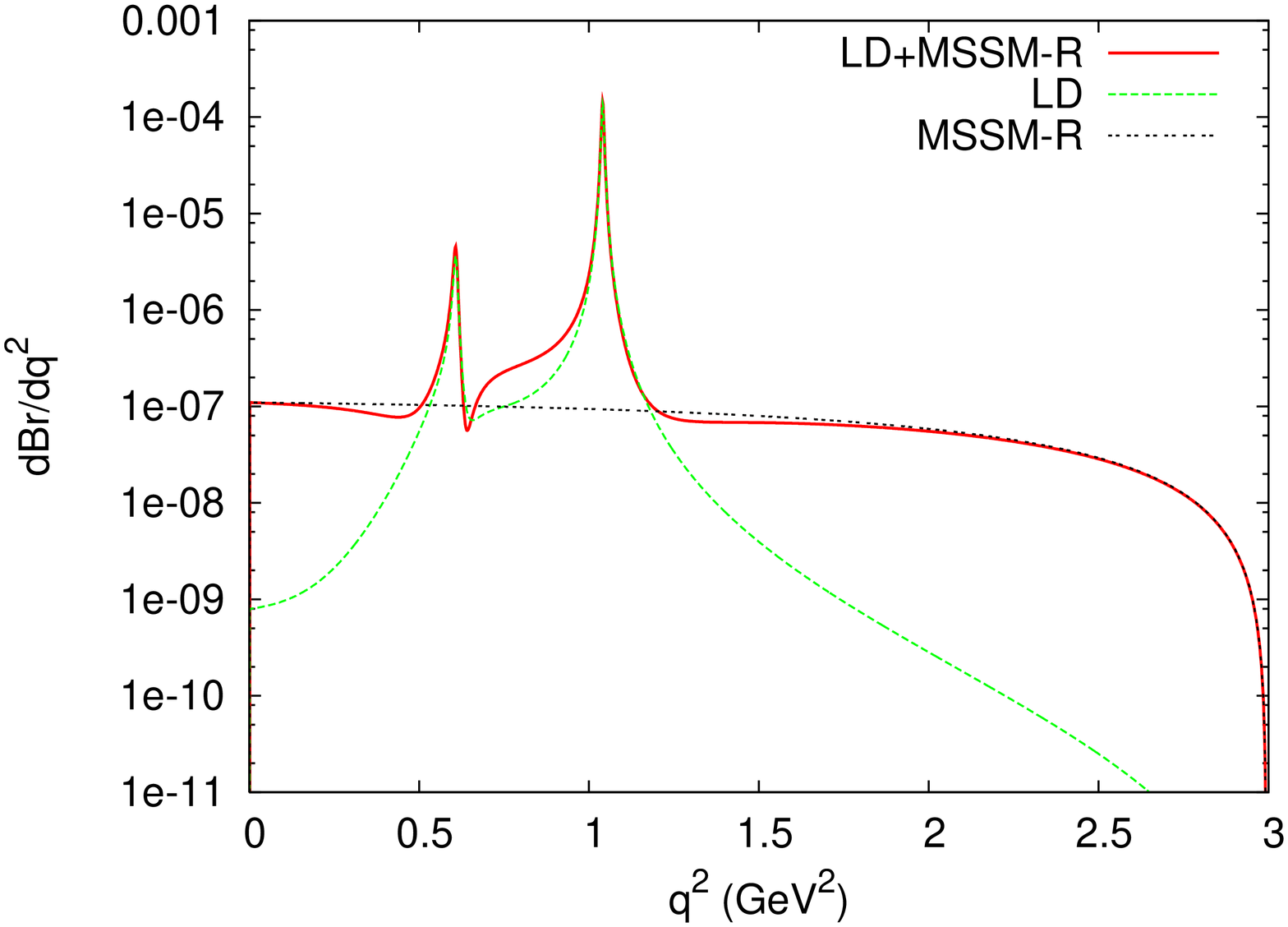}&
\includegraphics[angle=-90,width=0.5\textwidth]{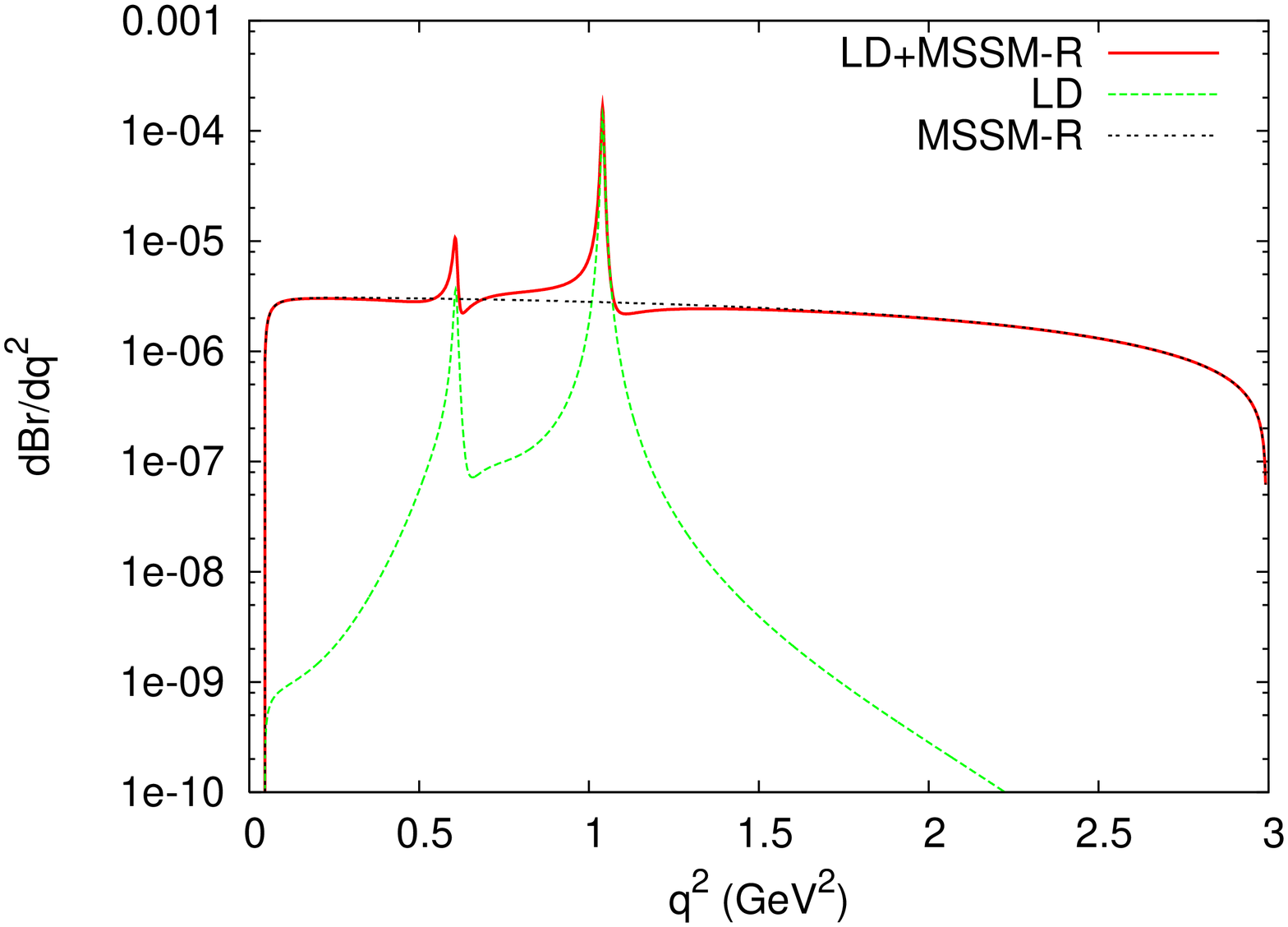}
\end{tabular}
\caption{Distributions of the maximal branching ratios in the
  MSSM$\slasha{R}$ model for the decay modes $D^+ \to \pi^+ e^+
  e^-$~(left) and $D^+ \to \pi^+ \mu^+ \mu^-$~(right). Full line
  represents the combined LD and SD contributions, and it corresponds
  to the experimental upper bound $BR(D^+ \to \pi^+ \mu^+ \mu^-) =
  8.8\E{-6}$ on the right plot.}
\label{fig:DRPV}
\end{figure}

On the other hand, the MSSM$\slasha{R}$ gives a slight increase to the
mode with electrons. Deviation from the LD amplitude is pronounced in
the region without resonances, where $m_{\ell\ell} < m_\rho$ or
$m_{\ell\ell} > m_\phi$ (Fig.~\ref{fig:DRPV}, left).
However, the most promising mode is the channel with muons. The
long-distance contribution ($2.2\E{-6}$) is a fair share of the
experimental upper bound~\cite{Dtopimumu} ($8.8\E{-6}$) and should be
taken into account together with the short distance part, when one is
constraining the Wilson coefficients. The difference is not big, i.e.
when we drop the LD part we get for the bound $|V_{cb}^* V_{ub}
C_{9,10}^\mu| < 27$, while the analysis with LD part included gives
\begin{equation} \label{eq:RPVbound}
  |V_{cb}^* V_{ub} C_{9,10}^\mu| < 23.
\end{equation}
The latter bound is a maximum with respect to the free relative phase
between the LD and MSSM$\slasha{R}$ parts of the amplitude.  Although
the inclusion of the LD term does not make substantial difference, it
will grow rapidly as experimental bound is approaching $2.2\E{-6}$.
All the branching ratios concerning MSSM$\slasha{R}$ and muons in the
final state~(Table~\ref{tab:Dbr},\ref{tab:Dsbr}) and their kinematical
distributions~(Fig.~\ref{fig:DRPV},\ref{fig:DsRPV}) use the
bound~(\ref{eq:RPVbound}).

\subsection{$D_s^+ \to K^+ \ell^+ \ell^-$}
\begin{figure}[!h]
\begin{tabular}{cc}
  \includegraphics[angle=-90,width=0.5\textwidth]{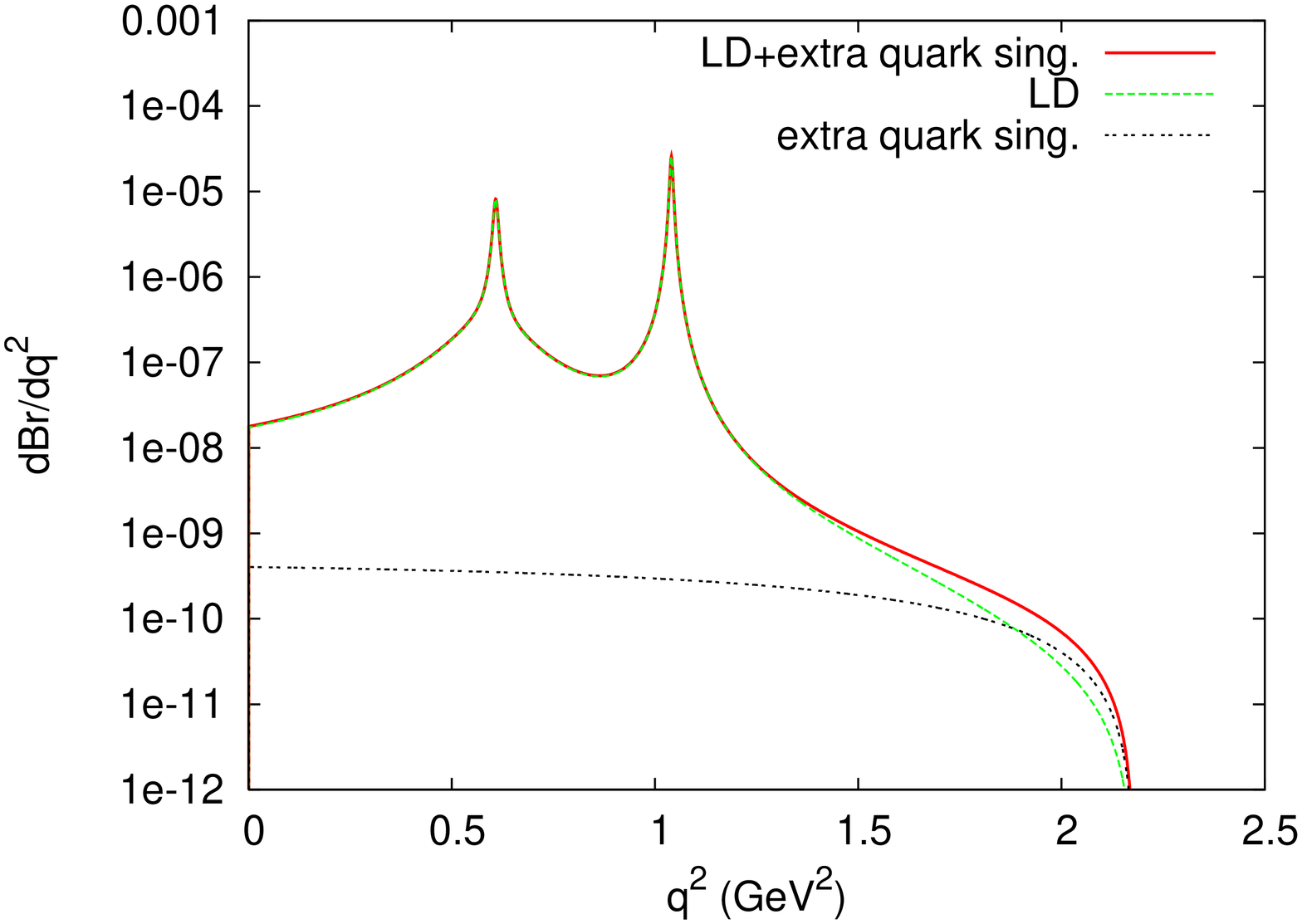} &
  \includegraphics[angle=-90,width=0.5\textwidth]{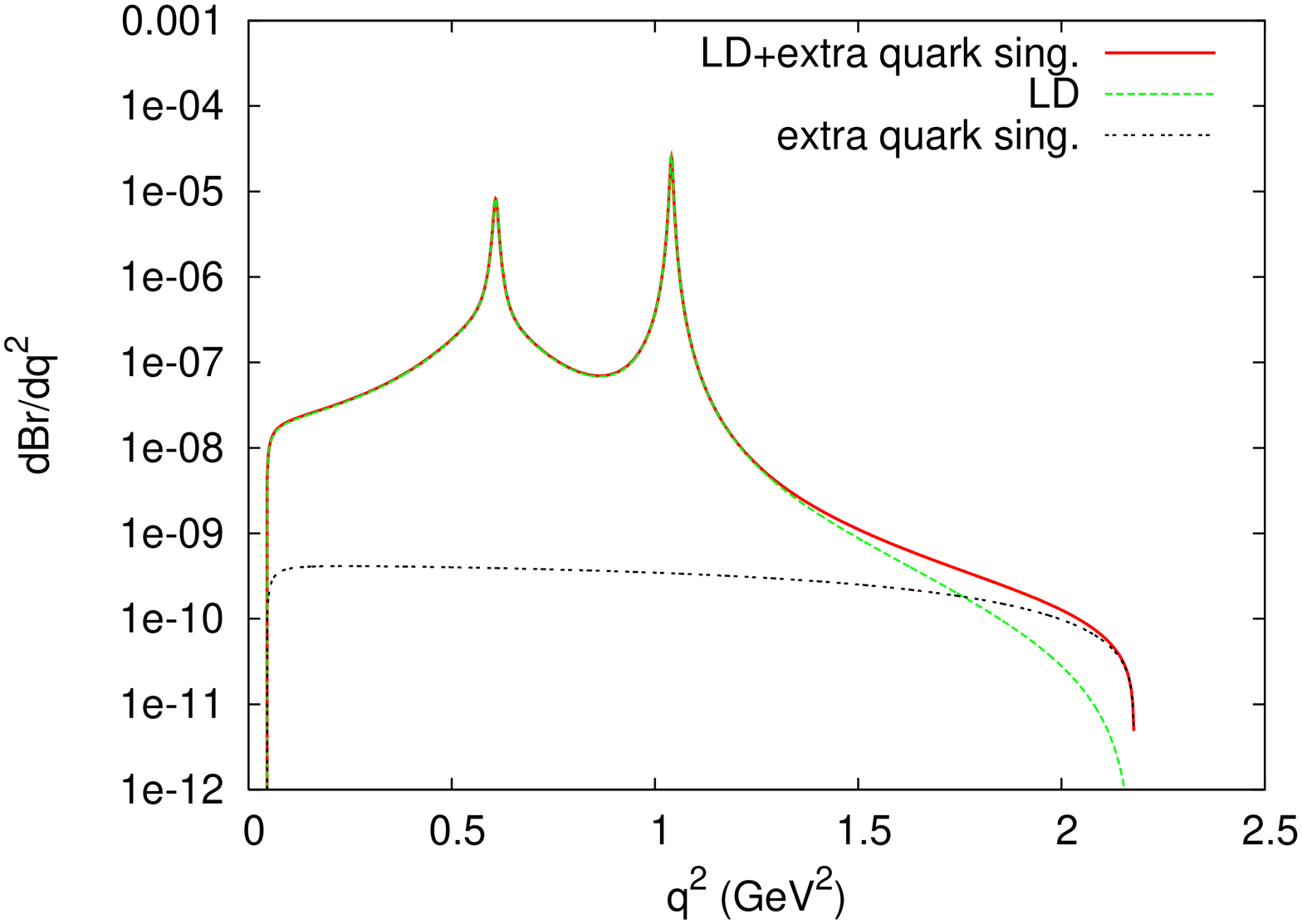}
\end{tabular}
\caption{Distributions of the maximal branching ratios in the model
  with extra quark singlet for the decay modes $D_s^+ \to K^+ e^+
  e^-$~(left) and $D_s^+ \to K^+ \mu^+ \mu^-$~(right). Full line
  represents the combined LD and SD contributions.}
\label{fig:DsEQS}
\end{figure}

\begin{figure}[!h]
\begin{tabular}{cc}
  \includegraphics[angle=-90,width=0.5\textwidth]{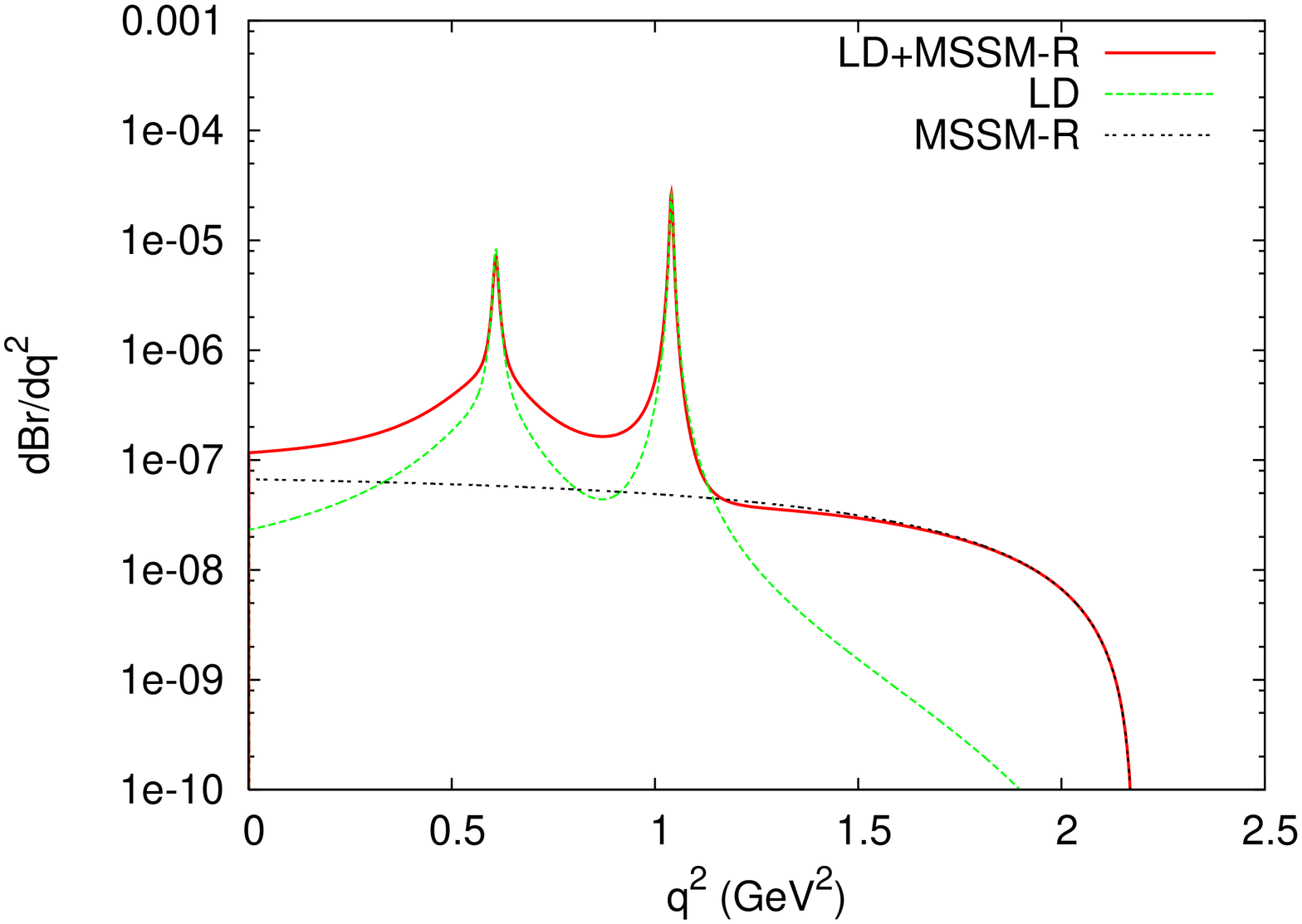} &
  \includegraphics[angle=-90,width=0.5\textwidth]{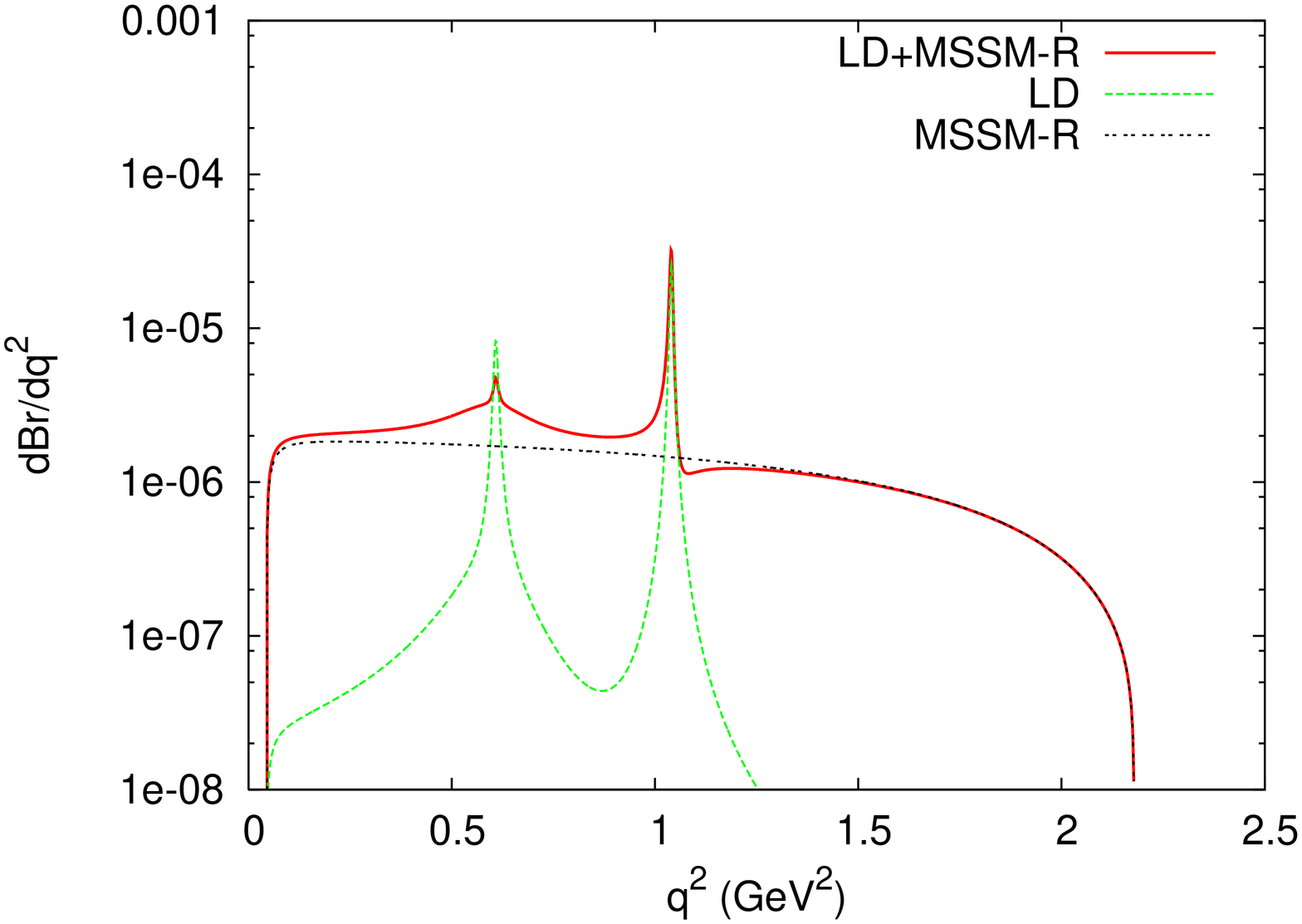}
\end{tabular}
\caption{Distributions of the maximal branching ratios in the
  MSSM$\slasha{R}$ model for the decay modes $D_s^+ \to K^+ e^+
  e^-$~(left) and $D_s^+ \to K^+ \mu^+ \mu^-$~(right).  Full line
  represents combined LD and SD contributions.}
\label{fig:DsRPV}
\end{figure}

The branching ratios contributions are summarized in
Table~\ref{tab:Dsbr}. Again, the EQS model has negligible effect
(Fig.~\ref{fig:DsEQS}). MSSM$\slasha{R}$ has a notable effect,
especially in the $\mu^+\mu^-$ mode, where it increases branching ratio by
an order of magnitude (Fig.~\ref{fig:DsRPV}). In this case, the
MSSM$\slasha{R}$ overshadows the LD contribution throughout the phase
space, except in the close vicinity of the LD resonant peaks.
\begin{table}[!h]
\begin{tabular}{c||c||c|c||c|c}
  mode & LD & EQS & LD+EQS & MSSM$\slasha{R}$ & LD+MSSM$\slasha{R}$\\\hline\hline
  $D_s^+ \to K^+ e^+ e^-$ & $6.0\E{-7}$ & $5.4\E{-10}$ & $6.0\E{-7}$ & $9\E{-8}$ & $7.6\E{-7}$\\\hline
  $D_s^+ \to K^+ \mu^+ \mu^-$ & $6.0\E{-7}$ & $6.2\E{-10}$ & $6.0\E{-7}$ & $2.6\E{-6}$ & $3.6\E{-6}$
\end{tabular}
\caption{Total branching fractions of the $D_s^+ \to K^+ \ell^+ \ell^-$ modes. In the first column (LD) are only long-distance BRs. The remaining four columns give maximal contributions of the SD physics models alone and also combined contributions of the SD and LD physics.}
\label{tab:Dsbr}
\end{table}

\section{SUMMARY}
Recently observed $D^0 - \bar D^0$ mass difference constrains the
value of tree-level flavor changing neutral coupling $c \to u Z$,
which is present in the models with an additional singlet up-like
quark.  We have studied the impact of this coupling on rare $D^+ \to
\pi^+ \ell^+ \ell^-$ and $D_s^+ \to K^+ \ell^+ \ell^-$ decays, where
its effects are accompanied by the long distance contributions. We
have determined long-distance contributions in $D_s^+ \to K^+ \ell^+
\ell^-$ following the same phenomenologically inspired model as it has
been done previously in the case of $D^+ \to \pi^+ \ell^+ \ell^-$. We
find that the effect of new extra singlet up-like quark is too small
to be seen in dilepton mass distributions for both decay modes.  In
our previous study we have considered forward-backward asymmetry in
the $D^0 \to \rho^0 \ell^+ \ell^-$ and found very small effect. New
constraint reduces that asymmetry even more, making it insignificant
for the experimental searches.

Present constraints on mass insertions in MSSM with conserved $R$-parity
still allow for increase of $c\to u\gamma$ rate by one order of
magnitude.  For the same reason MSSM could significantly increase
$c\to u\ell^+\ell^-$ rate at small $m_{\ell\ell}$. However, this MSSM
enhancement is not drastic in $D$ decays, since $D\to V\gamma$ and
$D\to V\ell^+\ell^-$ have large long distance contributions for small
$m_{\ell\ell}$, while $D\to P \ell^+\ell^-$ rate is multiplied by
factor of $m_{\ell\ell}^2$ due to gauge invariance.

The remaining possibility to search for new physics in rare $D$ decays
is offered by the MSSM models which contain R-parity violating terms.
We reinvestigate bounds on the combinations of these parameters in
$D^+ \to \pi^+ \ell^+ \ell^-$ by including the long-distance effects.
Using current upper bound on the rate for $D^+ \to \pi^+ \ell^+
\ell^-$ we derive new bound:
\begin{equation}
  \sum_{k=1}^3\left(\frac{100\e{GeV}}{M_{\tilde{d}^k_R}}\right)^2  \tilde{\lambda}'_{21k}   \tilde{\lambda}'_{22k} < 0.0041 
 \label{*}
\end{equation}
Since at Tevatron there are plans to investigate $D_s^+ \to K^+ \ell^+
\ell^-$ decay we use upper bound (\ref{*}) and calculate dilepton
invariant mass distribution. This bound still gives small increase of
the dilepton invariant mass distribution for the larger invariant
dilepton mass, making it attractive for the planned experimental
studies.

\begin{acknowledgments}
  This work is supported in part by the European Commission RTN
  network, Contract No. MRTN-CT-2006-035482 (FLAVIAnet), and by the
  Slovenian Research Agency.
\end{acknowledgments}

\bibliography{refs}

\begin{thebibliography}{45}
\expandafter\ifx\csname natexlab\endcsname\relax\def\natexlab#1{#1}\fi
\expandafter\ifx\csname bibnamefont\endcsname\relax
  \def\bibnamefont#1{#1}\fi
\expandafter\ifx\csname bibfnamefont\endcsname\relax
  \def\bibfnamefont#1{#1}\fi
\expandafter\ifx\csname citenamefont\endcsname\relax
  \def\citenamefont#1{#1}\fi
\expandafter\ifx\csname url\endcsname\relax
  \def\url#1{\texttt{#1}}\fi
\expandafter\ifx\csname urlprefix\endcsname\relax\def\urlprefix{URL }\fi
\providecommand{\bibinfo}[2]{#2}
\providecommand{\eprint}[2][]{\url{#2}}

\bibitem[{\citenamefont{Abe}(2007)}]{Belle0}
\bibinfo{author}{\bibfnamefont{K.}~\bibnamefont{Abe}}
  (\bibinfo{collaboration}{Belle}) (\bibinfo{year}{2007}),
  \eprint{hep-ex/0703036}.

\bibitem[{\citenamefont{Aubert et~al.}(2007)}]{BaBar0}
\bibinfo{author}{\bibfnamefont{B.}~\bibnamefont{Aubert}} \bibnamefont{et~al.}
  (\bibinfo{collaboration}{BABAR}) (\bibinfo{year}{2007}),
  \eprint{hep-ex/0703020}.

\bibitem[{\citenamefont{Schwartz}(2007)}]{schwartz}
\bibinfo{author}{\bibfnamefont{A.}~\bibnamefont{Schwartz}}
  (\bibinfo{year}{2007}), \bibinfo{note}{presented on 5th Flavor Physics and CP
  Violation Conference, Bled, Slovenia, 12-16 May 2007}.

\bibitem[{\citenamefont{Nir}(2007)}]{Nir0}
\bibinfo{author}{\bibfnamefont{Y.}~\bibnamefont{Nir}} (\bibinfo{year}{2007}),
  \eprint{hep-ph/0703235}.

\bibitem[{\citenamefont{Ciuchini et~al.}(2007)}]{IT0}
\bibinfo{author}{\bibfnamefont{M.}~\bibnamefont{Ciuchini}} \bibnamefont{et~al.}
  (\bibinfo{year}{2007}), \eprint{hep-ph/0703204}.

\bibitem[{\citenamefont{Blanke et~al.}(2007)\citenamefont{Blanke, Buras,
  Recksiegel, Tarantino, and Uhlig}}]{Buras0}
\bibinfo{author}{\bibfnamefont{M.}~\bibnamefont{Blanke}},
  \bibinfo{author}{\bibfnamefont{A.~J.} \bibnamefont{Buras}},
  \bibinfo{author}{\bibfnamefont{S.}~\bibnamefont{Recksiegel}},
  \bibinfo{author}{\bibfnamefont{C.}~\bibnamefont{Tarantino}},
  \bibnamefont{and} \bibinfo{author}{\bibfnamefont{S.}~\bibnamefont{Uhlig}}
  (\bibinfo{year}{2007}), \eprint{hep-ph/0703254}.

\bibitem[{\citenamefont{Chen et~al.}(2007)\citenamefont{Chen, Geng, and
  Yuan}}]{chen}
\bibinfo{author}{\bibfnamefont{C.-H.} \bibnamefont{Chen}},
  \bibinfo{author}{\bibfnamefont{C.-Q.} \bibnamefont{Geng}}, \bibnamefont{and}
  \bibinfo{author}{\bibfnamefont{T.-C.} \bibnamefont{Yuan}}
  (\bibinfo{year}{2007}), \eprint{arXiv:0704.0601 [hep-ph]}.

\bibitem[{\citenamefont{Li and Wei}(2007)}]{Li}
\bibinfo{author}{\bibfnamefont{X.-Q.} \bibnamefont{Li}} \bibnamefont{and}
  \bibinfo{author}{\bibfnamefont{Z.-T.} \bibnamefont{Wei}}
  (\bibinfo{year}{2007}), \eprint{arXiv:0705.1821 [hep-ph]}.

\bibitem[{\citenamefont{Golowich et~al.}(2007)\citenamefont{Golowich, Hewett,
  Pakvasa, and Petrov}}]{Golowich}
\bibinfo{author}{\bibfnamefont{E.}~\bibnamefont{Golowich}},
  \bibinfo{author}{\bibfnamefont{J.}~\bibnamefont{Hewett}},
  \bibinfo{author}{\bibfnamefont{S.}~\bibnamefont{Pakvasa}}, \bibnamefont{and}
  \bibinfo{author}{\bibfnamefont{A.~A.} \bibnamefont{Petrov}}
  (\bibinfo{year}{2007}), \eprint{arXiv:0705.3650 [hep-ph]}.

\bibitem[{\citenamefont{Burdman et~al.}(1995)\citenamefont{Burdman, Golowich,
  Hewett, and Pakvasa}}]{burdman1}
\bibinfo{author}{\bibfnamefont{G.}~\bibnamefont{Burdman}},
  \bibinfo{author}{\bibfnamefont{E.}~\bibnamefont{Golowich}},
  \bibinfo{author}{\bibfnamefont{J.~L.} \bibnamefont{Hewett}},
  \bibnamefont{and} \bibinfo{author}{\bibfnamefont{S.}~\bibnamefont{Pakvasa}},
  \bibinfo{journal}{Phys. Rev.} \textbf{\bibinfo{volume}{D52}},
  \bibinfo{pages}{6383} (\bibinfo{year}{1995}), \eprint{hep-ph/9502329}.

\bibitem[{\citenamefont{Burdman et~al.}(2002)\citenamefont{Burdman, Golowich,
  Hewett, and Pakvasa}}]{burdman2}
\bibinfo{author}{\bibfnamefont{G.}~\bibnamefont{Burdman}},
  \bibinfo{author}{\bibfnamefont{E.}~\bibnamefont{Golowich}},
  \bibinfo{author}{\bibfnamefont{J.}~\bibnamefont{Hewett}}, \bibnamefont{and}
  \bibinfo{author}{\bibfnamefont{S.}~\bibnamefont{Pakvasa}},
  \bibinfo{journal}{Phys. Rev.} \textbf{\bibinfo{volume}{D66}},
  \bibinfo{pages}{014009} (\bibinfo{year}{2002}), \eprint{hep-ph/0112235}.

\bibitem[{\citenamefont{Burdman and Shipsey}(2003)}]{burdman3}
\bibinfo{author}{\bibfnamefont{G.}~\bibnamefont{Burdman}} \bibnamefont{and}
  \bibinfo{author}{\bibfnamefont{I.}~\bibnamefont{Shipsey}},
  \bibinfo{journal}{Ann. Rev. Nucl. Part. Sci.} \textbf{\bibinfo{volume}{53}},
  \bibinfo{pages}{431} (\bibinfo{year}{2003}), \eprint{hep-ph/0310076}.

\bibitem[{\citenamefont{Fajfer and Prelovsek}(2006{\natexlab{a}})}]{FP-LH}
\bibinfo{author}{\bibfnamefont{S.}~\bibnamefont{Fajfer}} \bibnamefont{and}
  \bibinfo{author}{\bibfnamefont{S.}~\bibnamefont{Prelovsek}},
  \bibinfo{journal}{Phys. Rev.} \textbf{\bibinfo{volume}{D73}},
  \bibinfo{pages}{054026} (\bibinfo{year}{2006}{\natexlab{a}}),
  \eprint{hep-ph/0511048}.

\bibitem[{\citenamefont{Fajfer et~al.}(1999{\natexlab{a}})\citenamefont{Fajfer,
  Prelovsek, and Singer}}]{prelovsek1}
\bibinfo{author}{\bibfnamefont{S.}~\bibnamefont{Fajfer}},
  \bibinfo{author}{\bibfnamefont{S.}~\bibnamefont{Prelovsek}},
  \bibnamefont{and} \bibinfo{author}{\bibfnamefont{P.}~\bibnamefont{Singer}},
  \bibinfo{journal}{Eur. Phys. J.} \textbf{\bibinfo{volume}{C6}},
  \bibinfo{pages}{471} (\bibinfo{year}{1999}{\natexlab{a}}),
  \eprint{hep-ph/9801279}.

\bibitem[{\citenamefont{Fajfer et~al.}(1998)\citenamefont{Fajfer, Prelovsek,
  and Singer}}]{prelovsek2}
\bibinfo{author}{\bibfnamefont{S.}~\bibnamefont{Fajfer}},
  \bibinfo{author}{\bibfnamefont{S.}~\bibnamefont{Prelovsek}},
  \bibnamefont{and} \bibinfo{author}{\bibfnamefont{P.}~\bibnamefont{Singer}},
  \bibinfo{journal}{Phys. Rev.} \textbf{\bibinfo{volume}{D58}},
  \bibinfo{pages}{094038} (\bibinfo{year}{1998}), \eprint{hep-ph/9805461}.

\bibitem[{\citenamefont{Fajfer et~al.}(2001)\citenamefont{Fajfer, Prelovsek,
  and Singer}}]{prelovsek3}
\bibinfo{author}{\bibfnamefont{S.}~\bibnamefont{Fajfer}},
  \bibinfo{author}{\bibfnamefont{S.}~\bibnamefont{Prelovsek}},
  \bibnamefont{and} \bibinfo{author}{\bibfnamefont{P.}~\bibnamefont{Singer}},
  \bibinfo{journal}{Phys. Rev.} \textbf{\bibinfo{volume}{D64}},
  \bibinfo{pages}{114009} (\bibinfo{year}{2001}), \eprint{hep-ph/0106333}.

\bibitem[{\citenamefont{Bianco et~al.}(2003)\citenamefont{Bianco, Fabbri,
  Benson, and Bigi}}]{bigi0}
\bibinfo{author}{\bibfnamefont{S.}~\bibnamefont{Bianco}},
  \bibinfo{author}{\bibfnamefont{F.~L.} \bibnamefont{Fabbri}},
  \bibinfo{author}{\bibfnamefont{D.}~\bibnamefont{Benson}}, \bibnamefont{and}
  \bibinfo{author}{\bibfnamefont{I.}~\bibnamefont{Bigi}},
  \bibinfo{journal}{Riv. Nuovo Cim.} \textbf{\bibinfo{volume}{26N7}},
  \bibinfo{pages}{1} (\bibinfo{year}{2003}), \eprint{hep-ex/0309021}.

\bibitem[{\citenamefont{Prelovsek}(2000)}]{prelovsek0}
\bibinfo{author}{\bibfnamefont{S.}~\bibnamefont{Prelovsek}}
  (\bibinfo{year}{2000}), \eprint{hep-ph/0010106}.

\bibitem[{\citenamefont{Greub et~al.}(1996)\citenamefont{Greub, Hurth, Misiak,
  and Wyler}}]{greub}
\bibinfo{author}{\bibfnamefont{C.}~\bibnamefont{Greub}},
  \bibinfo{author}{\bibfnamefont{T.}~\bibnamefont{Hurth}},
  \bibinfo{author}{\bibfnamefont{M.}~\bibnamefont{Misiak}}, \bibnamefont{and}
  \bibinfo{author}{\bibfnamefont{D.}~\bibnamefont{Wyler}},
  \bibinfo{journal}{Phys. Lett.} \textbf{\bibinfo{volume}{B382}},
  \bibinfo{pages}{415} (\bibinfo{year}{1996}), \eprint{hep-ph/9603417}.

\bibitem[{\citenamefont{Bigi et~al.}(1990)\citenamefont{Bigi, Gabbiani, and
  Masiero}}]{masiero}
\bibinfo{author}{\bibfnamefont{I.~I.~Y.} \bibnamefont{Bigi}},
  \bibinfo{author}{\bibfnamefont{F.}~\bibnamefont{Gabbiani}}, \bibnamefont{and}
  \bibinfo{author}{\bibfnamefont{A.}~\bibnamefont{Masiero}},
  \bibinfo{journal}{Z. Phys.} \textbf{\bibinfo{volume}{C48}},
  \bibinfo{pages}{633} (\bibinfo{year}{1990}).

\bibitem[{\citenamefont{Prelovsek and Wyler}(2001{\natexlab{a}})}]{PW}
\bibinfo{author}{\bibfnamefont{S.}~\bibnamefont{Prelovsek}} \bibnamefont{and}
  \bibinfo{author}{\bibfnamefont{D.}~\bibnamefont{Wyler}},
  \bibinfo{journal}{Phys. Lett.} \textbf{\bibinfo{volume}{B500}},
  \bibinfo{pages}{304} (\bibinfo{year}{2001}{\natexlab{a}}),
  \eprint{hep-ph/0012116}.

\bibitem[{\citenamefont{Fajfer et~al.}(2003)\citenamefont{Fajfer, Singer, and
  Zupan}}]{jure}
\bibinfo{author}{\bibfnamefont{S.}~\bibnamefont{Fajfer}},
  \bibinfo{author}{\bibfnamefont{P.}~\bibnamefont{Singer}}, \bibnamefont{and}
  \bibinfo{author}{\bibfnamefont{J.}~\bibnamefont{Zupan}},
  \bibinfo{journal}{Eur. Phys. J.} \textbf{\bibinfo{volume}{C27}},
  \bibinfo{pages}{201} (\bibinfo{year}{2003}), \eprint{hep-ph/0209250}.

\bibitem[{\citenamefont{Prelovsek and Wyler}(2001{\natexlab{b}})}]{sasa}
\bibinfo{author}{\bibfnamefont{S.}~\bibnamefont{Prelovsek}} \bibnamefont{and}
  \bibinfo{author}{\bibfnamefont{D.}~\bibnamefont{Wyler}},
  \bibinfo{journal}{Phys. Lett.} \textbf{\bibinfo{volume}{B500}},
  \bibinfo{pages}{304} (\bibinfo{year}{2001}{\natexlab{b}}),
  \eprint{hep-ph/0012116}.

\bibitem[{\citenamefont{Fajfer and
  Prelovsek}(2006{\natexlab{b}})}]{faj-ichep06}
\bibinfo{author}{\bibfnamefont{S.}~\bibnamefont{Fajfer}} \bibnamefont{and}
  \bibinfo{author}{\bibfnamefont{S.}~\bibnamefont{Prelovsek}}
  (\bibinfo{year}{2006}{\natexlab{b}}), \eprint{hep-ph/0610032}.

\bibitem[{\citenamefont{Barger et~al.}(1995)\citenamefont{Barger, Berger, and
  Phillips}}]{barger}
\bibinfo{author}{\bibfnamefont{V.~D.} \bibnamefont{Barger}},
  \bibinfo{author}{\bibfnamefont{M.~S.} \bibnamefont{Berger}},
  \bibnamefont{and} \bibinfo{author}{\bibfnamefont{R.~J.~N.}
  \bibnamefont{Phillips}}, \bibinfo{journal}{Phys. Rev.}
  \textbf{\bibinfo{volume}{D52}}, \bibinfo{pages}{1663} (\bibinfo{year}{1995}),
  \eprint{hep-ph/9503204}.

\bibitem[{\citenamefont{Langacker and London}(1988)}]{lang}
\bibinfo{author}{\bibfnamefont{P.}~\bibnamefont{Langacker}} \bibnamefont{and}
  \bibinfo{author}{\bibfnamefont{D.}~\bibnamefont{London}},
  \bibinfo{journal}{Phys. Rev.} \textbf{\bibinfo{volume}{D38}},
  \bibinfo{pages}{886} (\bibinfo{year}{1988}).

\bibitem[{\citenamefont{Abel et~al.}(2003)\citenamefont{Abel, Masip, and
  Santiago}}]{abel}
\bibinfo{author}{\bibfnamefont{S.~A.} \bibnamefont{Abel}},
  \bibinfo{author}{\bibfnamefont{M.}~\bibnamefont{Masip}}, \bibnamefont{and}
  \bibinfo{author}{\bibfnamefont{J.}~\bibnamefont{Santiago}},
  \bibinfo{journal}{JHEP} \textbf{\bibinfo{volume}{04}}, \bibinfo{pages}{057}
  (\bibinfo{year}{2003}), \eprint{hep-ph/0303087}.

\bibitem[{\citenamefont{Higuchi and Yamamoto}(2000)}]{higuchi}
\bibinfo{author}{\bibfnamefont{K.}~\bibnamefont{Higuchi}} \bibnamefont{and}
  \bibinfo{author}{\bibfnamefont{K.}~\bibnamefont{Yamamoto}},
  \bibinfo{journal}{Phys. Rev.} \textbf{\bibinfo{volume}{D62}},
  \bibinfo{pages}{073005} (\bibinfo{year}{2000}), \eprint{hep-ph/0004065}.

\bibitem[{\citenamefont{Lee}(2004)}]{lee}
\bibinfo{author}{\bibfnamefont{J.~Y.} \bibnamefont{Lee}},
  \bibinfo{journal}{JHEP} \textbf{\bibinfo{volume}{12}}, \bibinfo{pages}{065}
  (\bibinfo{year}{2004}), \eprint{hep-ph/0408362}.

\bibitem[{\citenamefont{He et~al.}(2005)}]{CLEO_pll}
\bibinfo{author}{\bibfnamefont{Q.}~\bibnamefont{He}} \bibnamefont{et~al.}
  (\bibinfo{collaboration}{CLEO}), \bibinfo{journal}{Phys. Rev. Lett.}
  \textbf{\bibinfo{volume}{95}}, \bibinfo{pages}{221802}
  (\bibinfo{year}{2005}), \eprint{hep-ex/0508031}.

\bibitem[{\citenamefont{Freyberger et~al.}(1996)}]{CLEO_vll}
\bibinfo{author}{\bibfnamefont{A.}~\bibnamefont{Freyberger}}
  \bibnamefont{et~al.} (\bibinfo{collaboration}{CLEO}), \bibinfo{journal}{Phys.
  Rev. Lett.} \textbf{\bibinfo{volume}{76}}, \bibinfo{pages}{3065}
  (\bibinfo{year}{1996}).

\bibitem[{\citenamefont{Link et~al.}(2003)}]{FERMILAB_pll}
\bibinfo{author}{\bibfnamefont{J.~M.} \bibnamefont{Link}} \bibnamefont{et~al.}
  (\bibinfo{collaboration}{FOCUS}), \bibinfo{journal}{Phys. Lett.}
  \textbf{\bibinfo{volume}{B572}}, \bibinfo{pages}{21} (\bibinfo{year}{2003}),
  \eprint{hep-ex/0306049}.

\bibitem[{\citenamefont{Aitala et~al.}(2001)}]{FERMILAB_vll}
\bibinfo{author}{\bibfnamefont{E.~M.} \bibnamefont{Aitala}}
  \bibnamefont{et~al.} (\bibinfo{collaboration}{E791}), \bibinfo{journal}{Phys.
  Rev. Lett.} \textbf{\bibinfo{volume}{86}}, \bibinfo{pages}{3969}
  (\bibinfo{year}{2001}), \eprint{hep-ex/0011077}.

\bibitem[{\citenamefont{Casas and Dimopoulos}(1996)}]{CD}
\bibinfo{author}{\bibfnamefont{J.~A.} \bibnamefont{Casas}} \bibnamefont{and}
  \bibinfo{author}{\bibfnamefont{S.}~\bibnamefont{Dimopoulos}},
  \bibinfo{journal}{Phys. Lett.} \textbf{\bibinfo{volume}{B387}},
  \bibinfo{pages}{107} (\bibinfo{year}{1996}), \eprint{hep-ph/9606237}.

\bibitem[{\citenamefont{Fajfer et~al.}(1999{\natexlab{b}})\citenamefont{Fajfer,
  Prelovsek, and Singer}}]{FPS99}
\bibinfo{author}{\bibfnamefont{S.}~\bibnamefont{Fajfer}},
  \bibinfo{author}{\bibfnamefont{S.}~\bibnamefont{Prelovsek}},
  \bibnamefont{and} \bibinfo{author}{\bibfnamefont{P.}~\bibnamefont{Singer}},
  \bibinfo{journal}{Phys. Rev.} \textbf{\bibinfo{volume}{D59}},
  \bibinfo{pages}{114003} (\bibinfo{year}{1999}{\natexlab{b}}),
  \eprint{hep-ph/9901252}.

\bibitem[{\citenamefont{Fajfer and
  Kamenik}(2005{\natexlab{a}})}]{hep-ph/0412140}
\bibinfo{author}{\bibfnamefont{S.}~\bibnamefont{Fajfer}} \bibnamefont{and}
  \bibinfo{author}{\bibfnamefont{J.}~\bibnamefont{Kamenik}},
  \bibinfo{journal}{Phys. Rev.} \textbf{\bibinfo{volume}{D71}},
  \bibinfo{pages}{014020} (\bibinfo{year}{2005}{\natexlab{a}}),
  \eprint{hep-ph/0412140}.

\bibitem[{\citenamefont{Isgur and Wise}(1990)}]{IW}
\bibinfo{author}{\bibfnamefont{N.}~\bibnamefont{Isgur}} \bibnamefont{and}
  \bibinfo{author}{\bibfnamefont{M.~B.} \bibnamefont{Wise}},
  \bibinfo{journal}{Phys. Rev.} \textbf{\bibinfo{volume}{D42}},
  \bibinfo{pages}{2388} (\bibinfo{year}{1990}).

\bibitem[{\citenamefont{Becirevic et~al.}(2003)\citenamefont{Becirevic, Lubicz,
  Mescia, and Tarantino}}]{Becirevic}
\bibinfo{author}{\bibfnamefont{D.}~\bibnamefont{Becirevic}},
  \bibinfo{author}{\bibfnamefont{V.}~\bibnamefont{Lubicz}},
  \bibinfo{author}{\bibfnamefont{F.}~\bibnamefont{Mescia}}, \bibnamefont{and}
  \bibinfo{author}{\bibfnamefont{C.}~\bibnamefont{Tarantino}},
  \bibinfo{journal}{JHEP} \textbf{\bibinfo{volume}{05}}, \bibinfo{pages}{007}
  (\bibinfo{year}{2003}), \eprint{hep-lat/0301020}.

\bibitem[{\citenamefont{Jansen et~al.}(1996)}]{Jansen}
\bibinfo{author}{\bibfnamefont{K.}~\bibnamefont{Jansen}} \bibnamefont{et~al.},
  \bibinfo{journal}{Phys. Lett.} \textbf{\bibinfo{volume}{B372}},
  \bibinfo{pages}{275} (\bibinfo{year}{1996}), \eprint{hep-lat/9512009}.

\bibitem[{\citenamefont{Ho-Kim and Pham}(2000)}]{pham}
\bibinfo{author}{\bibfnamefont{Q.}~\bibnamefont{Ho-Kim}} \bibnamefont{and}
  \bibinfo{author}{\bibfnamefont{X.-Y.} \bibnamefont{Pham}},
  \bibinfo{journal}{Phys. Rev.} \textbf{\bibinfo{volume}{D61}},
  \bibinfo{pages}{013008} (\bibinfo{year}{2000}), \eprint{hep-ph/9906235}.

\bibitem[{\citenamefont{Allanach et~al.}(1999)\citenamefont{Allanach, Dedes,
  and Dreiner}}]{Allanach}
\bibinfo{author}{\bibfnamefont{B.~C.} \bibnamefont{Allanach}},
  \bibinfo{author}{\bibfnamefont{A.}~\bibnamefont{Dedes}}, \bibnamefont{and}
  \bibinfo{author}{\bibfnamefont{H.~K.} \bibnamefont{Dreiner}},
  \bibinfo{journal}{Phys. Rev.} \textbf{\bibinfo{volume}{D60}},
  \bibinfo{pages}{075014} (\bibinfo{year}{1999}), \eprint{hep-ph/9906209}.

\bibitem[{\citenamefont{Johns}(2002)}]{Dtopimumu}
\bibinfo{author}{\bibfnamefont{W.~E.} \bibnamefont{Johns}}
  (\bibinfo{collaboration}{FOCUS}) (\bibinfo{year}{2002}),
  \eprint{hep-ex/0207015}.

\bibitem[{\citenamefont{Lichard}(1999)}]{lichard}
\bibinfo{author}{\bibfnamefont{P.}~\bibnamefont{Lichard}},
  \bibinfo{journal}{Acta Phys. Slov.} \textbf{\bibinfo{volume}{49}},
  \bibinfo{pages}{215} (\bibinfo{year}{1999}), \eprint{hep-ph/9811493}.

\bibitem[{\citenamefont{Yao et~al.}(2006)}]{pdg}
\bibinfo{author}{\bibfnamefont{W.~M.} \bibnamefont{Yao}} \bibnamefont{et~al.}
  (\bibinfo{collaboration}{Particle Data Group}), \bibinfo{journal}{J. Phys.}
  \textbf{\bibinfo{volume}{G33}}, \bibinfo{pages}{1} (\bibinfo{year}{2006}).

\bibitem[{\citenamefont{Fajfer and Kamenik}(2005{\natexlab{b}})}]{kamenik}
\bibinfo{author}{\bibfnamefont{S.}~\bibnamefont{Fajfer}} \bibnamefont{and}
  \bibinfo{author}{\bibfnamefont{J.}~\bibnamefont{Kamenik}},
  \bibinfo{journal}{Phys. Rev.} \textbf{\bibinfo{volume}{D72}},
  \bibinfo{pages}{034029} (\bibinfo{year}{2005}{\natexlab{b}}),
  \eprint{hep-ph/0506051}.

\end{thebibliography}

\end{document}